\documentstyle[12pt,afterpage,epsfig,rotating]{article}
\begin{document}
\baselineskip = 23pt
\begin{center}
{\Large{\bf
Energy barriers for diffusion on stepped Pt(111) surface
}}
~\\
~\\
F. M\'{a}ca, M. Kotrla, {\em Institute of Physics, ASCR, Praha, Czech Republic}\\
O.S. Trushin, {\em Institute of Microelectronics, RASC, Yaroslavl, Russia}\\

\end{center}
~\\

\begin{abstract}
We performed molecular statics calculations of energy barriers
for adatom moves in the vicinity of steps on Pt(111) surface.
We used the semi-empirical many-body Rosato--Guillope--Legrand potential
and we systematically calculated barriers for descent of straight steps,
steps with a kink and  small islands as well as barriers for 
diffusion along the step edges.
We confirmed that 
the lowest barrier for descent  
is for an  exchange process 
near kink's or island's corner on a step with a \{111\} microfacet.
Diffusion along a step with a \{111\} microfacet is faster than
along a step with a \{100\} microfacet. We also calculated barriers
for diffusion on several surfaces vicinal to Pt(111).
Our results are compared with previous calculations.
\end{abstract}


\section{Introduction}
The diffusion in Pt/Pt(111) system 
has been recently subject of several studies
both experimental [1-3]
and theoretical 
[1,4-11]
It was motivated by growth experiments,
particularly by observation of the so-called reentrant layer-by-layer
growth \cite {kunkel90}
and also by growth-temperature-induced switch from a triangular growth shape
bounded by steps with a \{100\} microfacet to one bounded by steps with a \{111\} microfacet 
\cite{michely93}.
In order to explain these experiments on the microscopic level
a detailed understanding of the diffusion on inhomogeneous Pt(111)
surfaces is needed.

The self-diffusion energy barrier on  the flat Pt(111)
surface has been determined accurately in two independent experiments.
Bott et al. \cite{bott96} deduced  from 
scanning-tunneling-microscopy (STM) measurements of the density of 
platinum islands at different temperatures an activation energy
0.26$\pm$ 0.01 eV.
This is in remarkable agreement with
the recent field-ion-microscope (FIM) experiment  \cite{kyuno98}
giving the barrier 0.260$\pm$ 0.003 eV.
The agreement of  theoretical calculations
with the experimental value is rather 
unsatisfactory (see Table \ref{tab:flat}).
Semi-empirical potentials systematically underestimate the experimental value
and  the existing ab-initio full potential (FP) calculations give
in the contrary significantly higher energies.

Energy barriers for adatom moves on an inhomogeneous surface 
are not easily accessible by experiment and it is the
aim of the theory to calculate energy barriers for elementary
processes which are supposed to be 
relevant for explanation of growth experiments.
Several calculations have been already performed.
They confirmed that the layer-by-layer growth
at low temperatures can be explained by decrease of step-edge barrier
(called Ehrlich-Schwoebel barrier) for small and irregular islands
\cite{wang_r94,villarba94,s-liu96}.
As regards the shape of growing islands
it has been analyzed by several theoretical groups 
\cite{s-liu93,jacobsen95,jacobsen96},
but so far no mechanism has been agreed on.

\begin{table}[ht]
\vglue0mm
\begin{center}
\begin{tabular}{|l|l|c|c|}
\hline
~ &Method & Ref. & $E_S$   \\
\hline
\hline
~ & FIM & \cite{feibelman94} &    0.25$\pm$ 0.02\\
~ & STM & \cite{bott96}      &    0.26$\pm$ 0.01\\
\rule{0.5cm}{0pt}\begin{rotate}{90} {\large Exp.}\end{rotate} & FIM & \cite{kyuno98}    &    0.260$\pm$ 0.003\\
\hline
~ & Morse pot. & \cite{bassett78} & 0.07.\\
~ & \mbox{EAM} & \cite{liu91} & 0.078\\
~ & EAM & \cite{villarba94} & 0.08\\
~ & CEMT & \cite{wang_r94} & 0.038 \\
~ & EMT & \cite{stoltze94} & 0.16 \\
~ & EMT & \cite{s-liu94} & 0.13 \\
~ & EMT & \cite{jacobsen95} & 0.16 \\
\cline{2-4}
~ & FP & \cite{feibelman94} & 0.38\\
~ & FP & \cite{mortensen96} & 0.36\\
~ & FP & \cite{boisvert98} & 0.33\\
\cline{2-4}
\rule{0.5cm}{0pt}\begin{rotate}{90} {\large ~~~Theory}\end{rotate} & RGL & present & 0.17 \\
\hline
\end{tabular}
\caption{
Results for self-diffusion barrier $E_S$ (in eV) on flat Pt(111) surface, EAM - embedded atom
method, EMT - effective medium theory, FP - full potential calculations, for other abbreviations see
text.
\label{tab:flat}
}
\end{center}
\afterpage{\clearpage}
\end{table}

In the situation when there is not quantitative agreement between theory
and experiment already for diffusion on the flat surface,
 one can still look at trends and compare mutually values
of energy barriers for various adatom moves obtained within one approach.
However, different authors investigated various processes
using different potentials and the comparison is not always possible.
Therefore, it is desirable to compare  barriers for
as many as possible processes using one approach.

In this paper we use
 the semi-empirical many-body Rosato--Guillope--Legrand (RGL) potential
\cite{rosato89,cleri93} developed within the second moment
approximation of tight-binding theory.
This potential allows relatively fast calculations
and, at the same time, provides reasonably reliable results 
\cite{trushin97a}.
We show that with the RGL potential we can reproduce majority of
previous results.
In particular, we systematically study exchange processes
for descent of steps and diffusion along straight step edges.

\section{Method}
The simulations were done with finite atomic slabs with a free surface
on the top, two atomic layers fixed on the bottom, and periodic boundary
conditions in the two directions parallel to the surface. The slab
representing the substrate of (111) surface was 11 layers thick with 448
atoms per layer. For diffusion along channels on the vicinal surfaces (311),
(211), (331), (221) and (322) we used the systems of approximately 5000 
atoms that contained 19 to 44 layers, with 110 to 240 atoms per layer.

The equations of motion were solved using a leap-frog algorithm
\cite{frog} with a time step of~$0.5\cdot 10^{-14}$~s. We used the classical
{\em NVE} ensemble and molecular-dynamics cooling method for molecular
statics calculations of energy barriers. A conventional
spherical cut-off with the cutoff radius of $6.4$~\AA ~was used.
We have applied the RGL potentials with 
the parameters obtained by a fit to experimental data
taking into account interactions up to the fifth nearest neighbors
\cite{cleri93}.

The  energy barrier of a particular diffusion process was obtained by
testing systematically various paths of adatom moves, and the path
with the lowest energy barrier was chosen to be the optimum path. The
adatom diffusion barrier $E_d$ is defined as $E_d = E_{sad} - E_{min}$
where $E_{sad}$ and $E_{min}$ are the total energy of the system with
the adatom at the saddle point and at the equilibrium adsorption site,
respectively. The minimum energy paths for both the jump and exchange
processes were calculated. The minimum energy path for jump diffusion
was determined by allowing the migrating atom to relax in a plane
perpendicular to the path at each step. The rest of the atoms in the
system was allowed to relax in all directions. The energy barrier of
the exchange process was obtained by moving the surface atom, which
was to be replaced, by an adatom with finite steps along the direction
of exchange. This atom was allowed to relax in the plane perpendicular
to the exchange direction at each step, whereas the other atoms,
including the adatom, were allowed to relax in all directions.

\section{Results}
\subsection{Descent to the lower terrace}
As the first step we calculated the energy barrier for diffusion on the flat
Pt(111) surface. We found the value 0.17 eV, which is comparable with 
the values obtained from the effective medium theory (see Table \ref{tab:flat}).

We studied descent of an adatom to the lower terrace 
from both types of steps  on the (111) surface, 
i.e. , step A with \{100\} microfacet 
and step B with \{111\} microfacet (see Fig. 1).
We performed calculations for several different geometries:
straight steps, steps with a kink, island 3 $\times$ 3
and  island 6 $\times$ 6.
We considered both direct jumps of an adatom and exchange processes.
For all considered geometries we systematically studied
all possible adatom moves, and  investigated the dependence of
energy barriers on the local geometry.
Our results are summarized and compared with previous calculations
in Table \ref{tab:des}.

\begin{table}[hb]
\vglue0mm
\begin{center}
\begin{tabular}{|l|l|c|c|c|c|}
\hline
Step & Process& RGL &EMT \cite{s-liu96} & EMT \cite{jacobsen95} & 
EAM \cite{villarba94}\\
\hline
\hline
A & Jump over straight step                  &0.53&0.46&0.41&0.7-0.8 \\
 & Jump over kink                   &0.49&0.41& -  & - \\
 & Exchange over straight step              &0.50&0.56& -  &0.25 \\
 & Exchange over kink (4 $\rightarrow$ rf2)           &0.45& - &  - & -   \\
 & Exchange over kink (5 $\rightarrow$ rf3)            &0.42& 0.48  &  - & -   \\
 & Exchange over kink (4 $\rightarrow$ rf3)          &0.48& -  &  - &0.08 \\
  & Exchange next to kink (8 $\rightarrow$ rf4)         &0.50&0.55& -  &  -    \\
  & Exchange next to corner (2 $\rightarrow$ rf1)       &0.47&-   & -  &-      \\
\hline
B   & Jump over straight step                  &0.52&0.46&- &0.7-0.8 \\
 & Jump over kink                   &0.49&0.41& -  & - \\
 & Exchange over straight step              &0.40&0.31&0.37 &0.18 \\
 & Exchange over kink (4 $\rightarrow$ rf2)          & 0.42  &  - & - & - \\
 & Exchange over kink (5 $\rightarrow$ rf3)            &0.44& -  &  - & 0.2   \\
 & Exchange over kink (6 $\rightarrow$ rf3)           &0.45&0.46&  - & -   \\
  & Exchange next to kink (8 $\rightarrow$ rf4)         &0.39&0.30& -  &  -    \\
  & Exchange next to corner (2 $\rightarrow$ rf1)       &0.26&-   &0.26  &0.06      \\

\hline
\end{tabular}
\caption{
Energy barriers  (in eV) for descent of steps
on Pt(111). For the meaning of labels see Fig. 2.
\label{tab:des}
}
\end{center}
\afterpage{\clearpage}
\end{table}

Let us consider at first the case of straight steps.
The energy barrier for direct jump from the upper to the lower terrace is
0.53 eV for step A and 0.52  eV 
for step B.
The barrier for exchange at the step A (0.50 eV)
is comparable with the barrier for jump,
but at the step B, the barrier for exchange (0.40 eV) is significantly lower.
We did not find a difference in binding energy
between fcc and hcp site on the flat surface. 
However, we observed a difference between  binding energies at
fcc and hcp positions near the steps on the upper terrace.
We found that near the step A 
the fcc site is energetically favored by
0.02 eV in comparison with the hcp site,
but  on step B  the hcp site
is energetically favored by the same value
in comparison with the fcc site.

\begin{table}[h]
\begin{center}
\begin{tabular}{|l|c|c|c|}
\hline
~ & No. & Exchange &  Energy (eV)\\
\hline
\hline
~ & 1 & 1 $\rightarrow$ rf1 & 0.32 \\
\cline{2-4}
~ & 2 & 2 $\rightarrow$ rf1 &  { 0.26} \\
\cline{2-4}
\rule{0.5cm}{0pt}\begin{rotate}{90} {\Large  rf1}\end{rotate} &
3 & 3 $\rightarrow$ rf1 &  0.29 \\
\hline
~ & 4 & 3 $\rightarrow$ rf2 & 0.43 \\
\cline{2-4}
\rule{0.5cm}{0pt}\begin{rotate}{90} {\Large  rf2}\end{rotate} &
5 & 4 $\rightarrow$ rf2 &  { 0.42} \\
\hline
~ & 6 & 4 $\rightarrow$ rf3 & 0.72 \\
\cline{2-4}
~ & 7 & 5 $\rightarrow$ rf3 & {0.44} \\
\cline{2-4}
~ & 8 & 6 $\rightarrow$ rf3 & 0.45  \\
\cline{2-4}
\rule{0.5cm}{0pt}\begin{rotate}{90} {\Large rf3}\end{rotate}
& 9  & 7 $\rightarrow$ rf3 &  0.46 \\
\hline
~ & 10 & 7 $\rightarrow$ rf4 & {0.39} \\
\cline{2-4}
~ & 11 & 8 $\rightarrow$ rf4 & 0.39 \\
\cline{2-4}
\rule{0.5cm}{0pt}\begin{rotate}{90} {\Large rf4}\end{rotate} &
12 & 9 $\rightarrow$ rf4 & 0.40 \\
\hline
\end{tabular}
\end{center}
~\\
\caption{
Summary of energy barriers for exchange on a kink on step B.
\label{tab:stepb}}
\afterpage{\clearpage}
\end{table}

In geometry with a kink on a step one has to consider more processes.
As an example of our approach we show in Table \ref{tab:stepb}
the energy barriers for different exchange processes near a kink
on step B.
We calculated energy barriers for all possible exchange events
and for each exchange process all moves of the replaced atom were considered.
For example in the case of the exchange process presented in the first line
of Table \ref{tab:stepb} (the adatom starts in the fcc site labeled by {\bf 1}
and pushes  out the edge atom {\bf rf1}) two
possible different paths schematically shown by arrows in Fig. 4 were considered.
Only the minimal
barrier heights are presented in Table \ref{tab:stepb}.

There is only small influence of kink geometry on energy barrier for direct
jump (Tab. \ref{tab:des}).
Contrary to Liu and Metiu \cite{s-liu96} but in agreement with
Villarba et al. \cite{villarba94} and Jacobsen et al. \cite{jacobsen95}
we found 
that the diffusion barrier for exchange on steps with kinks is significantly
lower than on straight steps.
We found that the smallest energy barrier for exchange on the step with a kink
is on the step A for exchange
over the kink atom (exchange of adatom from the hcp position {\bf 5} 
with the edge atom
{\bf rf3} - see Fig. 2) and on the step B for exchange {\em next 
to corner} (from the hcp
position {\bf 2} to the edge atom position {\bf rf1}). 

We continued the investigation of geometry effects on energy barriers
for descent
with the study of small islands ($3\times 3$ and $6 \times 6$ atoms) on Pt(111)
surface.
We found no changes in the barrier heights for direct jumps.
For exchange processes on small island further reduction of energy barriers has
been observed. For $3\times 3$ island minimal values were obtained for the exchange of the atom
in the middle of the edge (0.26 eV for A-type edge and 0.13 eV
for B-type edge).
For the 36-atom island
the minimal energy barriers (0.30 eV for A-type edge and 0.16 eV for
B-type edge) were found for
the exchange processes starting in the hcp sites in the vicinity of the
island's corner moving away the edge neighbors of the corner atoms.
The exchange
mechanism is more important for diffusion on small islands than for diffusion in the
vicinity of longer atomic edge.
\\                         

\subsection{Diffusion along the step edges}

We also calculated barriers for diffusion along straight steps 
(Fig. 3).
We found that diffusion along the step of type A is slower
(the barrier is 0.55 eV) than along the step of type B
(the barrier is 0.43 eV) - see Fig. 3.
This is in agreement with several previous calculations 
(0.44 eV and 0.40 eV) \cite{s-liu93},
(0.60 eV and 0.43 eV) \cite{villarba94},  (0.23 eV and 0.18 eV) 
\cite{jacobsen95}. But it is in contradiction with the assumption made 
in the recent kinetic Monte Carlo simulation \cite{jacobsen96}
attempting to explain
island shapes observed in homo-epitaxial growth of Pt(111).
The authors justified their assumption
by comparison with the results of field-ion-microscope measurements
on  the (311) and (331) surfaces \cite{bassett78,kellogg86}.
These surfaces consist of narrow terraces 
with step edges of type  A and B, respectively.
Experimental values of energy barriers are $E_{\rm 311}=0.60\pm 0.03$ eV 
\cite{kellogg86}, and  $E_{\rm 311}=0.84\pm 0.1$ eV \cite{bassett78}.
Nevertheless, it may be misleading to use barriers obtained for 
the (311) and (331) surfaces instead of barriers along the step edges 
on the (111) surface.
It is apparent (see Fig. 4) 
that when not only nearest neighbors,
but also more distant neighborhood is considered, then the geometry is different 
and one can expect different contributions of adatom-surface interactions
to the energy barriers.
The relaxation of vicinal surfaces is also different.
In principle one cannot exclude even qualitative difference
as it seems to be the case for platinum and as it has been observed on
iridium \cite{fu96}.
Unfortunately up to our best knowledge, 
there is no experimental result for diffusion 
along step edges on Pt(111).

\begin{table}[htb]
\vglue0mm
\begin{center}
\begin{tabular}{|c|c|c|c|c|c|}
\hline
Surface & Morse pot.\cite{bassett78} & EAM-VC \cite{liu91}& 
EAM-AFW\cite{liu91} & EAM \cite{villarba94} & 
Exp. \cite{bassett78,kellogg86}\\
\hline
\hline
311 & 0.49 & 0.63 & 0.43 & 0.57 & 0.60 $\pm$ 0.03\\
\hline 
331 & 0.71 & 0.54 & 0.28 & 0.53 & 0.84 $\pm$ 0.1 \\
\hline 
\end{tabular}
\caption{Comparison of activation energy barriers (in eV) for
diffusion on the (311) and (331) surfaces.
\label{tab:311}
}
\end{center}
\afterpage{\clearpage}
\end{table}

Table \ref{tab:311} shows results of previous calculations for diffusion
along steps on the  (311) and (331) surfaces.
Three EAM calculations in Table \ref{tab:311}
differ by parameterization.
We can see that all EAM calculations give a lower 
barrier on the (331) surface than on the (311) surface which is
in contradiction with the experiment.
Only the calculation using the Morse potential gives 
qualitative agreement with the experiment.

In our calculation using the RGL potential, we obtained 
 the same barriers  0.50 eV  for
diffusion along steps on both (311) and (331)
surfaces.
We calculated also barriers for diffusion along steps on some other vicinal
surfaces  with wider terraces: (211), (221) and (332) shown in Fig. 4.
All our results are summarized in Table \ref{tab:along}.
There is a clear tendency with the changing 
distance between steps.
In the case of surface geometries with A-type steps the energy barrier is increasing with the step-step distance whereas for the B-type steps
it is decreasing.
We note that 
the same trend is valid also for the data reported in Ref. \cite{villarba94}.

\begin{table}[htb]
\vglue0mm
\begin{center}
\begin{tabular}{|c|c|c|c|}
\hline
\multicolumn{2}{|c|}{ Step A} & 
\multicolumn{2}{c|}{ Step B}\\
\hline
Surface & $E_d$ &Surface & $E_d$\\
\hline
111 & 0.55 & 111 & 0.43 \\
211 & 0.53 & 332 & 0.45 \\
311 & 0.50 & 221 & 0.48 \\
- & - & 331 & 0.50 \\
\hline 
\end{tabular}
\end{center}
\caption{Activation energy barriers $E_d$ (in eV) for
diffusion along steps on different surfaces calculated with 
the RGL potential.
\label{tab:along}}
\afterpage{\clearpage}
\end{table}

\section{Conclusion}
We systematically calculated barriers for descent of straight steps,
steps with a kink and  small islands, and we determined the 
processes with a minimal energy barrier.
We have shown that with the many-body RGL potential 
we can obtain results comparable with
those obtained from the effective medium theory.
We found that the descent 
is easier at the step B than at the step A and
that for irregular geometry the exchange mechanism of diffusion must be taken systematically into
account. The lowest energy barrier for steps with kinks is significantly lower than for 
straight steps and the
energy barrier for descent from small islands is even much lower.

In the case of diffusion along step edges
we obtained lower barriers
for  diffusion along
the step B  than for diffusion along the step A.
Calculations for vicinal surfaces demonstrate
the importance of the step-step interaction on the energy barriers.
We observed the trend that the diffusion along 
the step B on Pt(111) surface is faster than along the step A.

\vspace{5mm}
\begin{center}

{\large\bf Acknowledgment}\\
\end{center}   
 Financial support for this work was provided by the Grant Agency
of the Czech Republic (Project 202/96/1736).

\bibliography{../../my_bib/journals,../../my_bib/md,../../my_bib/exper,../../my_bib/simgr,../../my_bib/kr,../../my_bib/gen}

\newpage
~\\
{\Large \bf Figure captions}\\
~\\
Fig. 1. Two types of step edges, A and B, for a large island. Solid line
shows the diffusion path along the island edge. The atomic layers from
the surface to the bulk are large filled circles, large open circles,
small open circles, and very small open circles.
\label{fig:along} \\
~\\
Fig. 2. Different exchange processes near a
 kink site on the step B on Pt(111) surface.
The edge atoms undergoing investigated exchange diffusion (rf1,...,rf4) 
and starting positions of an adatom
(1,...,9) are shown.
\label{fig:stepB}\\
~\\
Fig. 3. The dependence of adatom energy for the path along the 
edge of a large island shown in Fig. 1.
\label{fig:energy}\\
~\\
Fig. 4. Surface normal views for Pt surfaces where self-diffusion
along the A-step type edge, i.e. (311) and (211), and along the B-step type
edge, i.e. (331), (221) and (332), has been calculated. Solid line
shows  the diffusion path along the edge for each surface. The smaller
filled circles are, the deeper the atomic layers lie in
the bulk. \\
\label{fig:vicinals}
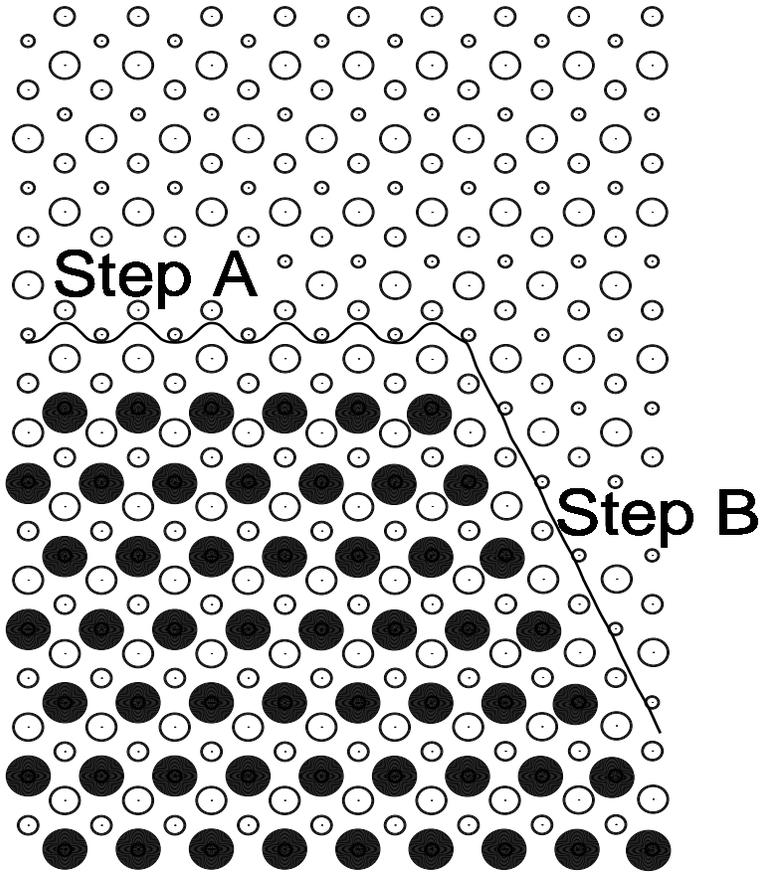
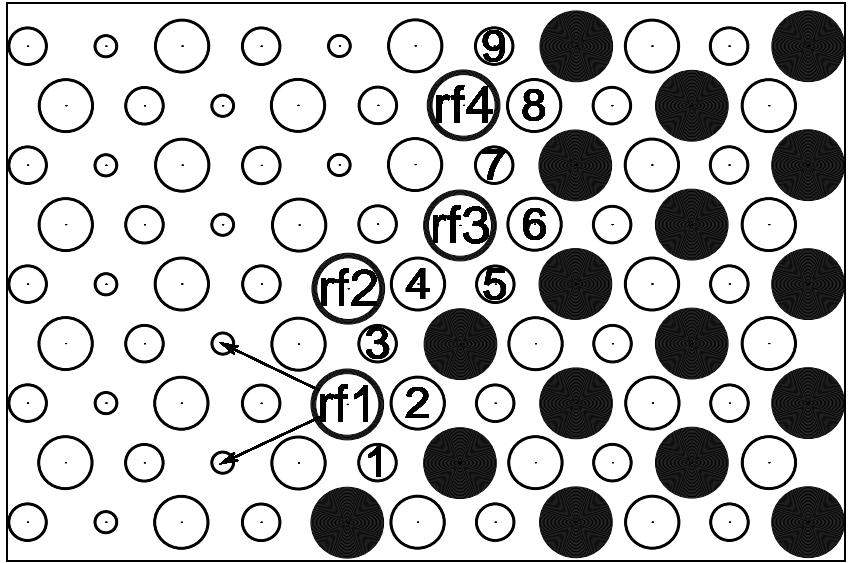
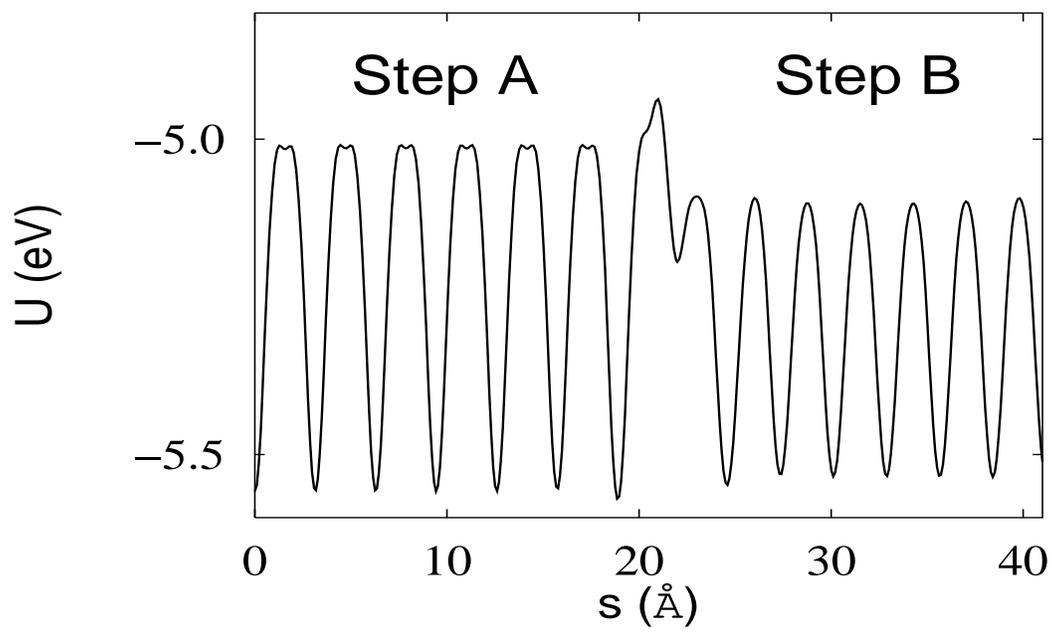
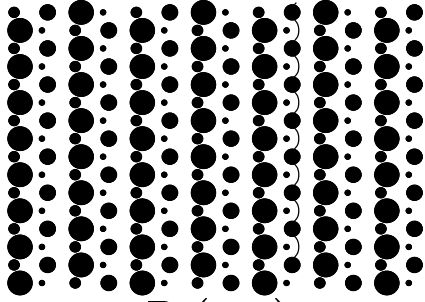
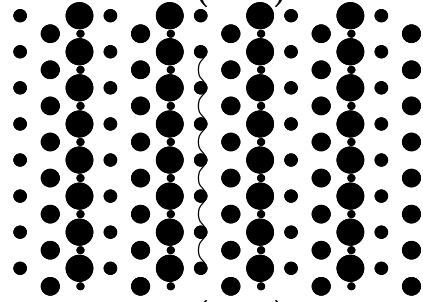
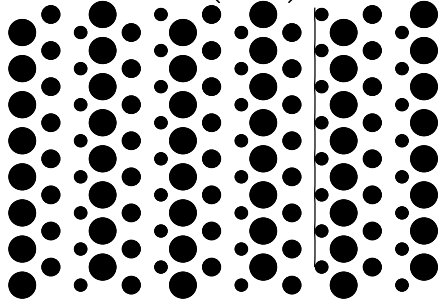
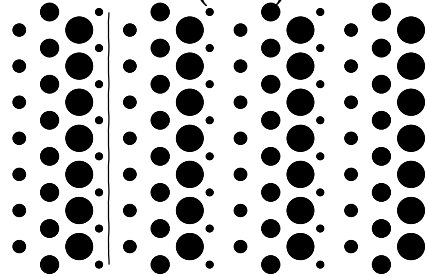
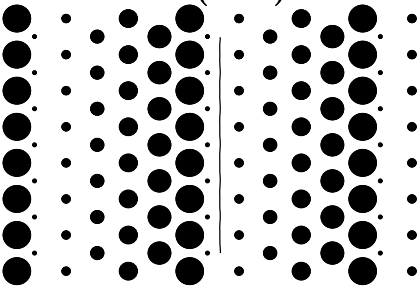
\end{document}